\documentclass[12pt]{iopart}
\usepackage{graphicx}
\usepackage{url}
\usepackage{color}
\begin{document}
\jl{2}

\title{Investigating the nuclear Schiff moment of $^{207}$Pb in ferroelectric PbTiO$_3$}

\author{J. A. Ludlow$\footnote{Now at AquaQ Analytics,
Forsyth House, Cromac Square, Belfast, United Kingdom, BT2 8LA}$ and O. P. Sushkov
}

\address{School of Physics, University of New South Wales, Sydney 
2052, Australia}

\begin{abstract}
A positive experimental measurement of the nuclear Schiff moment
would have important implications for physics beyond the standard model.
To aid in the interpretation of a proposed experiment to measure
the nuclear Schiff moment of $^{207}$Pb in the ferroelectric PbTiO$_3$,
three-dimensional Hartree-Fock calculations have been performed to model
the local electronic structure in the vicinity of the Pb nucleus.
The energy shift due to the Schiff moment is found to be 
a factor of 2 smaller in comparison to existing estimates.
\end{abstract}


\maketitle

\section{Introduction}

The search for the permanent electric-dipole moment (EDM) of quantum particles
has been of continued interest for over 40 years since the discovery of the
violation of the combined symmetry of charge conjugation ($C$) and parity ($P$)    
in the decay of the $K^0$ meson \cite{meson}. By the $CPT$ theorem, the existence of $CP$-violation
also implies that time-reversal ($T$) symmetry is violated \cite{gross}. $P$-violation, 
together with $T$-reversal asymmetry gives rise to a permanent 
electric-dipole moment of a quantum system in a stationary state. Therefore ongoing
searches for EDM's of elementary particles, nuclei, atoms and molecules are important
for studies of fundamental symmetries \cite{khrip} and they provide important constraints on
theories that attempt to go beyond the standard model \cite{he}.

Currently, no EDM has been experimentally observed. Experiments on paramagnetic
atoms provide the best upper limit on the electron EDM, with the most stringent
limitation coming from experiments with atomic thallium \cite{thal} and the YbF molecule \cite{Hudson}
\begin{eqnarray}
{\rm Tl}:\ \ \ &&d_e < 1.6\times10^{-27}e\,{\rm cm}.\nonumber\\
{\rm YbF}:\ \ \ &&d_e < 1.05\times10^{-27}e\,{\rm cm} \ .
\end{eqnarray} 
For diamagnetic atoms, the most EDM sensitive experiment has been performed with mercury vapour, $^{199}{\rm Hg}$, giving
an upper limit for the EDM of the $^{199}{\rm Hg}$ atom of \cite{griffith},
\begin{equation}
d(^{199}{\rm Hg}) \leq 3.1\times10^{-29}e\,{\rm cm}.
\end{equation} 
For diamagnetic atoms the major contribution
to the electron dipole moment is from the nuclear Schiff moment (NSM), {\bf S}. The NSM can be defined by the P- and T-odd electrostatic 
potential \cite{nsm},
\begin{equation}
\label{SD}
\varphi({\bf r})=4\pi ({\bf S\cdot\nabla})\delta({\bf r}).
\end{equation} 
The result of \cite{griffith} has been interpreted with the aid of calculations \cite{nsm1,nsm2} to
yield an upper limit of the Schiff moment of the $^{199}$Hg nucleus of,
\begin{equation}
S(^{199}{\rm Hg}) < 0.8\times 10^{-26} e\, a_B^3,
\end{equation} 
where $a_B$ is the Bohr radius.

Currently, there is intense interest in exploring physical systems that potentially
can provide orders of magnitude increases in sensitivity to CP violating effects.
As first discussed by Shapiro \cite{shapiro}, condensed matter systems are 
particularly promising in this regard. The electron's EDM is aligned with the spin of the electron, 
Aand therefore its magnetic moment. It follows that 
in a compound with uncompensated spins, the application of an external electric field will align
the EDM's and thus align the magnetic moments, giving rise to a macroscopic magnetization.
Experiments then aim to measure the reversal of the magnetization by reversing the direction of 
the external electric field. Recent efforts have focused on gadolinium 
garnets \cite{lam,sush1,buhmann,lam1} and more recently on the 
ceramic Eu$_{0.5}$Ba$_{0.5}$TiO$_3$ \cite{eur2plus10,nature,eur2plus,eckel}.

Similarly to the atomic case, condensed matter systems without unpaired electrons should also
be sensitive to the NSM. Ferroelectric PbTiO$_3$ has a huge effective internal electric field 
and therefore it has been suggected for searches of parity and time reversal symmetry 
violations \cite{legget}. The material was proposed for measuring the NSM in 
Ref. \cite{sush} where the first estimate of the expected effect for also performed. 
Experiments with this compound offer the possibilty of an improvement in sensitivity to the 
NSM by several orders of magnitude in comparison with existing 
results \cite{sush,phdthesis,budker,ludlow,pbrelax}. 

There has been a recent suggestion to detect axion dark matter with cold molecules \cite{Graham}.
It is possible that usage of PbTiO$_3$  as the dark matter detector has advantages compared to
molecules \cite{dark}. On the theoretical side the solid state part of the calculation of the axion 
sensitivity is exactly similar to that of the NSM. Actually the axion effect can be directly recalculated
from  the NSM. Here, we will only consider the NSM, but it is important to note the importance of this
problem to axions as well.

The NSM can be detected experimentally either by a measurement of the macroscopic magnetization induced 
by an electric field or by nuclear magnetic resonance. 
The lead nuclei will interact with the large internal electric field
in the ferroelectric leading to a P, T-odd energy shift. Here, we perform three-dimensional 
Hartree-Fock calculations in order to model the local electronic structure of PbTiO$_3$  
around the Pb nucleus and determine the energy shift caused by the NSM of the $^{207}$Pb nucleus.

\section{Model calculations}

PbTiO$_3$ is an ionic crystal made up of Pb$^{2+}$, Ti$^{4+}$ and O$^{2-}$ ions.
In the ferroelectric phase, PbTiO$_3$ is tetragonal with $c/a=1.065$, $a=3.902 \mbox{\AA}$,
$c=4.156 \mbox{\AA}$ \cite{nelmes}, with Pb$^{2+}$ and Ti$^{4+}$ displacements of $0.47 \mbox{\AA}$
and $0.30 \mbox{\AA}$ with respect to their non-ferroelectric positions \cite{warren}.
The ion displacements are in accordance with the Schiff theorem \cite{schiff}. The 
lattice relaxes so as to screen the strong internal ferroelectric field such that the
average field that acts on each charged particle in the compound is zero. 
However, the finite size of the ions plays an important role, with oxygen electrons ($s$ and $p$-wave)  
penetrating into the core of the Pb ion creating, due to the Pb displacement, a gradient of 
electron density at the Pb nucleus. 
The NSM interacts with this gradient, leading to an energy shift.
The electron density gradient can be expressed in terms of the mixing
of s- and p-electron orbitals at the Pb nucleus. In the semiclassical approximation
this leads to the following formula for the NSM energy shift $\Delta\epsilon$ \cite{sush1},
\begin{equation}\label{deleps}
\Delta\epsilon/E_0\simeq 
\frac{16\beta}{\sqrt{3}} \frac{Z^2}{(\nu_s\nu_p)^{3/2}}
\left( \frac{1}{3}R_{1/2}+\frac{2}{3}R_{3/2} \right)
\frac{({\bf X\cdot S})}{a_Bea_B^3}.
\end{equation}
Here ${\bf X}$ is the ferroelectric Pb ion displacement with respect to the
Oxygen coordination sphere, $E_0=27.2$ eV, $e=|e|$ is the 
charge of the electron, $Z=82$ is the nuclear charge of Pb, 
$\nu_s$ and $\nu_p$ are effective principal quantum numbers and 
  $\beta$ is a parameter describing lead orbitals 
partially occupied by oxygen electrons, see below.
$R_{1/2}$ and $R_{3/2}$ are relativistic enhancement factors \cite{khrip,sush1} 
given by,
\begin{eqnarray}\label{enhan}
R_{1/2}&=&\frac{4\gamma_{1/2}x_0^{2\gamma_{1/2}-2}}{[\Gamma(2\gamma_{1/2}+1)]^2}\nonumber\\
R_{3/2}&=&\frac{48\gamma_{1/2}x_0^{\gamma_{1/2}+\gamma_{3/2}-3}}{\Gamma(2\gamma_{1/2}+1)\Gamma(2\gamma_{3/2}+1)},
\end{eqnarray}
where $\gamma_{1/2}=\sqrt{1-Z^2\alpha^2}$, $\gamma_{3/2}=\sqrt{4-Z^2\alpha^2}$, $\alpha$ is the
fine structure constant, $\Gamma(x)$ is the gamma function and $x_0=(2Zr_0/a_B)$, with $r_0$ the nuclear radius.
Note that in the present work we follow the definition of the 
NSM in Eq. (\ref{SD}) that differs by a factor of $4\pi$ from that used in Ref. \cite{sush1}.

An understanding of the order of magnitude of the energy shift (\ref{deleps})
can be gained by considering that the energy shift can be approximated by the product 
of the internal electric
field in ferroelectric PbTiO$_3$, $E_{int}\simeq 10^8$ V/cm, and the EDM of the Pb$^{2+}$ 
ion, $d({\rm Pb}^{2+})/(e\, cm)\simeq 10^{-2}S/(ea_B^3)$, where we have assumed that 
the EDM of the Pb$^{2+}$ ion is similar to that of the Hg atom. This gives,
$\Delta\epsilon\simeq E_{int}d({\rm Pb}^{2+})\simeq 10^{6}\frac{S}{ea_B^3}\,{\rm eV}$ 
\cite{phdthesis}.

In Ref. \cite{sush}, Eq. (\ref{deleps}) was applied to calculate the energy shift
using a value of $\beta=-0.29$ estimated previously for GdO$_8$ clusters \cite{sush1}.
This gave an estimate for the energy shift of,
\begin{equation}\label{enshift}
\Delta\epsilon\simeq -1.1\times 10^6 \frac{S}{ea_B^3}\,{\rm eV}.
\end{equation}
The method used to estimate $\beta$ involved matching a linear combination
of the electron wavefunctions of the $2p_{\sigma}$ electrons of O$^{2-}$ 
possessing the correct cubic symmetry, with single particle $6s$ and $6p$
states of the central Gd$^{3+}$ ion that can be thought of as being occupied by the oxygen electrons. 
The matching is done at an intermediate distance in the cluster between the O$^{2-}$ ions and the 
central Gd$^{3+}$ ion.

In the present paper we perform a much more accurate calculation as compared to
the estimate of \cite{sush}.
It is well known that relativistic effects are very important for the calculation of the NSM.
Therefore in an ideal case one should calculate the many-body relativistic electron wave function
of the material and then calculate the expectation value of the operator (\ref{SD}) with
this wave function. Certainly in this case one has to use the relativistic version of
the operator (\ref{SD}) which is known \cite{nsm}.
There are the following complexities in such a calculation (i) the multi-center nature of the
electron wave function, (ii) electron correlations effects, (iii) relativistic effects.
Clearly the ideal calculation is impossible, one needs to do simplifications.

Our first simplification concerns point (iii), relativistic effects. This simplification is 
based on experience with the calculation of the NSM in atoms. Since the NSM interaction is a local
operator the relativistic effects can be factorized into relativistic factors as is done
in Eq. (\ref{deleps}). So, we will use Eq. (\ref{deleps}) instead of (\ref{SD}). This assumes
that the many-body electron wave function is factorized at the nucleus. The factorization
is certainly valid in the Hartree-Fock approximation, however, it is much better than that.
It is known that all Brueckner type correlations can be re-absorbed in the single particle
orbitals, the wave function remains factorized. 
Thus, we will use the non-relativistic Hamiltonian,
\begin{equation}
\label{H}
H=\sum_{electrons}\frac{{\bf p}_i^2}{2m}+\sum_{electrons}\frac{e^2}{|{\bf r}_i-{\bf r}_j|}
-\sum_{electrons,  nuclei}\frac{e^2Z_n}{|{\bf r}_i-{\bf R}_n|} 
\end{equation}
in our calculations and account for relativistic effects via Eq. (\ref{deleps}).
Based on experience with atoms we believe that the account of relativistic
effects via this procedure does not bring an inaccuracy
higher than 10\%, probably less. To confirm the accuracy of the account of relativistic effects in
terms of relativistic factors introduced in the non-relativistic
calculation one can compare the exact relativistic calculation of parity
violation in the Cs atom \cite{csrel} with effective non-relativistic
calculations (with relativistic factors) \cite{Khriplovich,Khripcs}. The difference is roughly 4\%, 
with the major part of the difference arising from 
correlations accounted for in \cite{csrel} and omitted in \cite{Khriplovich,Khripcs}.

The second simplification concerns point (ii), correlation effects. We disregard
correlation effects and use the Hartree-Fock method. This simplification bothers us much more
than relativistic effects. Even in atoms correlation effects can give up to $\sim$ 50\%
correction to matrix elements of local operators.
However, matrix elements of the NSM for systems without unpaired electrons are always more
stable than say matrix elements of the usual weak interaction, see, e.g. Refs.
\cite{nsm1,Khriplovich,bakasov,nsm2}.
We hope that the inaccuracy due to unaccounted correlation effects does not exceed 30\%.
Direct account of the correlation effects computationally is hardly possible in this system.
A feasible way to estimate the effect of correlations is to perform  a similar calculation 
using the DFT method. The DFT effectively accounts for Brueckner correlations, and therefore
a comparison between Hartree-Fock and DFT results would allow an estimate of the correction
due to correlations. In the present work we perform the Hartree-Fock calculation.

\begin{figure}[ht]
\begin{center}
\includegraphics[clip,width=10cm]{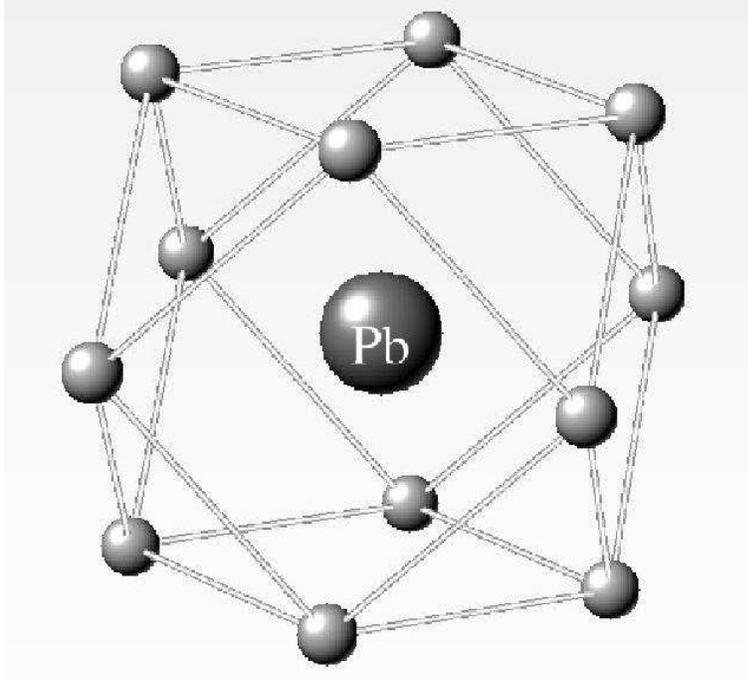}
\end{center}
\caption{The Pb$^{2+}$ ion surrounded by the first coordination sphere of 12 
O$^{2-}$ ions.}
\label{pbo12}
\end{figure}

The last simplification is about point (i), the multi-center nature of the problem.
We cannot do calculations for the infinite crystal and hence replace the crystal by the first
coordination sphere, the PbO$_{12}$ cluster shown in  Fig. \ref{pbo12}.
An {\it Ab-initio} Hartree-Fock calculation is performed for this  cluster.
The 12 O$^{2-}$ ions are placed at the mid-points of the edges of a cube
of length $4 \mbox{\AA}$. So the radius of the first coordination sphere
is approximately $2\sqrt{2}\mbox{\AA}$. The Pb$^{2+}$ ion is displaced by a distance
$X\sim 0.5\mbox{\AA}$ from the centre of the cube, this is the ferroelectric displacement.
The dynamic part of the cluster is surrounded by static charges.
The 8 Titanium ions are positioned approximately at the corners of the cube, this is the second
coordination sphere around Pb with the radius $2 \sqrt{3}\mbox{\AA}$.
The Ti ions are displaced from the corners of the cube by $0.6 X$.
Here we take into account that the Ti ferroelectric displacement is 0.6 of the Pb displacement  \cite{warren}.
We do not include the Ti ions explicitly in the Hartree-Fock calculation, instead we replace them by 
positive pointlike charges $q$. In our calculations we vary the effective Ti charge in the range $+3 < q < +4$.
Finally we place negative pointlike charges at the positions of the 24 O$^{2-}$ ions surrounding 
the PbO$_{12}$ cluster. This is the third coordination sphere with radius $2 \sqrt{6}\mbox{\AA}$.
The value of each pointlike negative charge also depends on $q$ and is tuned to ensure that the 
system is neutral. We do not explicitly include the Pb$^{2+}$ ions which are at the even smaller distance of
$4\mbox{\AA}$, however implicitly the Pb ions are also included since the neutrality condition is
ensured and since the precise radius of the third coordination sphere is not very important.
PbTiO$_3$ is a very good insulator with an electronic gap of about $4$ eV.
This justifies the cluster model. However, we will see that the NSM effect somewhat depends on the value of 
$q$ which we use. Therefore, it would be desirable to extend the size of the dynamic cluster to
PbO$_{12}$Ti$_8$ or even larger. Unfortunately, we were not able to do so because of lack of
convergence of the Hartree-Fock procedure. To estimate the inaccuracy of the present cluster
truncation we study the dependence of the NSM effect on $q$, see below. Based on this dependence we 
estimate the inaccuracy at the level of 15\%.

So the balance of our worries (inaccuracies) is the following. Correlation effects $\sim 30\%$,
cluster size $\sim 15\%$, relativistic effects less than 10\%.
Overall we estimate the inaccuracy at the level of 30-50\%.

The GAMESS-US quantum chemistry code \cite{gamess} is used to perform
Hartree-Fock calculations for PbO$_{12}$. The WTBS 
basis \cite{wtbs,wtbs1} set is used for all calculations presented here.
A diffuse sp shell is also added to the Pb$^{2+}$ and O$^{2-}$ orbitals in order to better represent 
the ionic bonds present in this system. A trial wavefunction is built from Pb$^{2+}$ and 
O$^{2-}$ Hartree-Fock wavefunctions, where in both cases the ions are enclosed by a cubical 
array of point charges. For O$^{2-}$ this is necessary in order to obtain a stable ground state 
in the Hartree-Fock approximation. This trial wavefunction is then used as a starting
approximation to obtain a converged Hartree-Fock wavefunction for the PbO$_{12}$ cluster.
To check that the WTBS basis is sufficient for our purposes we first performed atomic
calculations using this basis. By this, we mean atomic calculations of energy levels and NSM matrix
elements for isolated Pb$^{2+}$, Pb$^{1+}$, ions and for the isolated Oxygen atom.
Results of the calculations were compared with known atomic results obtained by different methods.
Parameters of the WTBS basis were adjusted to provide the best possible agreement. This is 
especially important to describe correctly the Pb interior. All in all the inaccuracy related to the 
WTBS basis can hardly exceed 1\% and is completely negligible compared to the balance of errors presented above.
Using  converged Hartree-Fock wavefunctions of the PbO$_{12}$ cluster we calculate the electron density
distribution around the Pb nucleus and then evaluate the NSM energy shift (\ref{deleps})
using the procedure described below.

 $\Delta\rho(z)$ is defined as the electron density asymmetry around the lead nucleus due to
the displacement of the lead ion from the center of the PbO$_{12}$ cluster,
$z$ is the distance from the lead nucleus in the direction of the
displacement. This can be represented by, 
\begin{equation}\label{delden}
\Delta\rho(z)=\frac{\rho(z)-\rho(-z)}{2}.
\end{equation}
On the one hand this asymmetry is directly calculated via the Hartree-Fock procedure
and on the other hand it can be re-expressed in terms of Pb atomic orbitals.
 Near the Pb$^{2+}$ ion  the electrostatic potential for electrons is dominated
by that of the central Pb$^{2+}$ ion. Therefore the 
wavefunctions of electrons in this region can be well described by partially occupied Pb$^{2+}$ orbitals.
The interaction of electrons with the NSM of Pb is local, therefore only $s-$ and $p$-waves
contribute to the matrix element.
First consider our cluster without ferroelectric displacement, $X=0$.
Let $|S\rangle$ and $|P\rangle$ be orbitals of the PbO$_{12}$ cluster
that have $s-$ and $p-$wave symmetries with respect to the 
centre of the cluster. These orbitals must be proportional to atomic
7s and 6p orbitals of the central lead ion.
\begin{equation}
|S\rangle\rightarrow\beta_S|7s\rangle,\,\,\,\,\,
|P_i\rangle\rightarrow\beta_P|6p_i\rangle, 
\end{equation}
where $i=x,y,z$.
Note that the choice of 7s and 6p orbitals of Pb is arbitrary, all external
orbitals behave similarly close to Pb nucleus. For example one can chose
8s orbital instead of 7s and redefine $\beta$,  $\beta_s|7s\rangle={\overline{\beta_s}}|8s\rangle$.
When the Pb ion is shifted from the centre of the
cluster, the $|S\rangle$ and $|P\rangle$ states no longer have 
exact $s$ and $p-$wave symmetries with respect to the lead ion 
but are instead a mixture of the two. 
For a small ferroelectric displacement ${\bf X}_i$ one always can use the linear approximation
\begin{eqnarray}
|S\rangle&\rightarrow&\beta_s|7s\rangle+\sum_i \beta^{\prime}_s\frac{X_i}{a_B}
|6p_i\rangle\nonumber\\
|P_i\rangle&\rightarrow&\beta_p|6p_i\rangle+
\sum_i \beta^{\prime}_p\frac{X_i}{a_B}|7s\rangle. 
\end{eqnarray}
The change in the density due to the displacement of the Pb$^{2+}$ ion along the
$z$ direction is then,
\begin{equation}\label{delta}
\Delta\rho(z)=4\beta\frac{X}{a_B}\psi_{7s}(z)\psi_{6p}(z),
\end{equation}
where $\beta=(\beta_s\beta^{\prime}_s+\beta_p\beta^{\prime}_p)$ and a factor of 2
comes from the double occupancy of each orbital.
Lead atomic orbitals 7s and 6p are known from atomic calculations. Therefore 
comparing (\ref{delta}) with (\ref{delden}) we find the coefficient $\beta$
and hence we calculate the NSM matrix element  (\ref{deleps}).
Actually we even do not need to know the effective principal quantum numbers
$\nu_s$ and $\nu_p$ since the matrix element  (\ref{deleps}) can be rewritten
directly in terms of $\psi_{7s}(r\to 0)$ and $\psi_{6p}(r\to 0)$ which are known
numerically.

\section{Results}

In order to test the various assumptions made in the present model, a number of 
calculations have been carried out. Firstly, the point charges representing the Ti ions are
treated here as an adjustable parameter in order to assess the dependence of the final result on the
crystal structure, with calculations carried out for a range of Ti charges.  
Secondly, the assumed linearity of the density shift is tested by carrying out calculations for a 
number of Pb displacements.

\begin{figure}[ht]
\begin{center}
\includegraphics[clip,width=10cm]{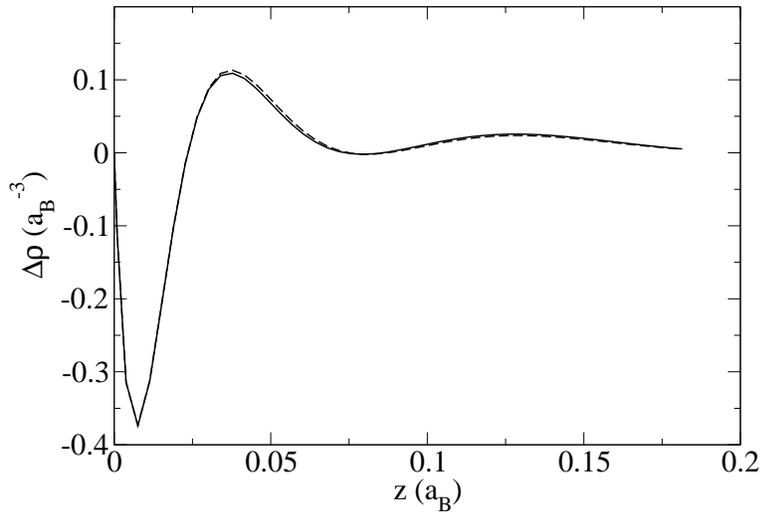}
\end{center}
\caption{$\Delta\rho(z)$ for X=0.5\AA, $q=+3.5$. Solid line, 
GAMESS-US; dashed line, Eq. (\ref{delta}) with $\beta=-0.125$.}
\label{deleho3.5}
\end{figure}

First, to illustrate how well Eq. (\ref{delta}) performs in 
capturing the essential physics of the density shift, 
$\Delta\rho$ as a function of the distance from the Pb nucleus
is plotted in Figure \ref{deleho3.5} for a Ti charge of +3.5.
We match the numerical density shift to the analytical 
expression of Eq. (\ref{delta}) using atomic Hartree-Fock wavefunctions for 
the $6p$ and $7s$ states of Pb$^{2+}$, treating $\beta$ as an
adjustable parameter. The GAMESS-US density is not reliable at very small distances because of
limitations of the WTBS basis. Therefore, the matching is done at points 
above the first node  at $r > 0.02a_B$, and
then an average value of $\beta$ is taken. As is clear, the analytical expression
agrees well with the numerical results. The solid GAMESS-US line is practically
indistinguishable from the dashed line given by Eq. (\ref{delta}).

\begin{figure}[ht]
\begin{center}
\includegraphics[clip,width=10cm]{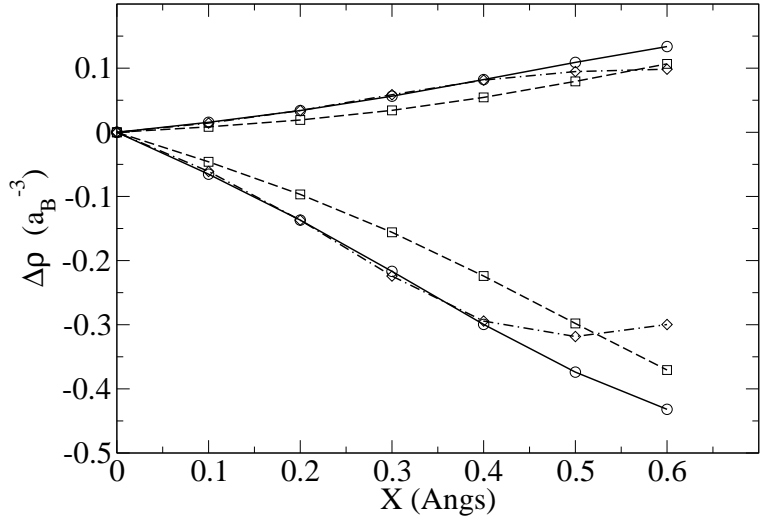}
\end{center}
\caption{Density shift as a function of the Pb ion displacement $X$,
(with Ti charges displaced by $0.3X/0.5$) for varying Ti charges $q$. 
Dashed line, $q=+3.0$; solid line, $q=+3.5$, dot-dashed, $q=+4.0$.
Lower lines at $z=0.008a_B$, upper lines at $z=0.04a_B$.}
\label{varti}
\end{figure}

Next, the validity of the linear approximation for the density shift is investigated.
Figure \ref{varti} plots $\Delta\rho$ at distances from the Pb nucleus corresponding
to the first minimum and maximum of $\Delta\rho$, as a function of the Pb$^{2+}$ ion 
displacement. From the figure, it is seen that a Ti charge of 3.5 gives the best
agreement with the linear approximation out to $X=0.5$\AA, with some deviation
from linearity observed for charges of 3.0 and 4.0. Most likely these small deviations 
observed only at very small $z$ are due to limitations of the WTBS basis. In any case 
the deviations are insignificant.

\begin{figure}[ht!]
\begin{center}
\includegraphics[clip,width=10cm]{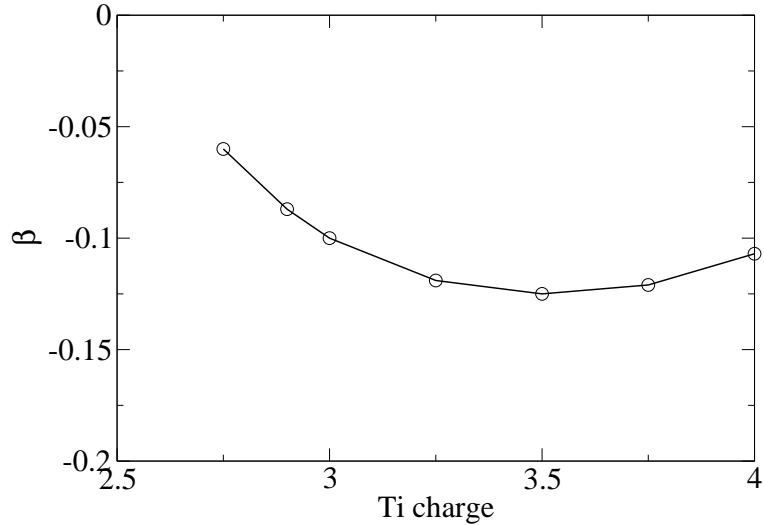}
\end{center}
\caption{$\beta$ as the Ti effective charge $q$ is varied for $X=0.5$\AA.}
\label{betaplot}
\end{figure}

Finally, $\beta$ as a function of the Ti charge is plotted in Figure \ref{betaplot}
for a Pb$^{2+}$ displacement of $X=0.5$\AA.
It is seen that for Ti charges between +3 and +4, $\beta$ ranges from -0.1 to -0.125, demonstrating 
that $\beta$ does not depend strongly on the crystal structure. This compares to a previous estimate of 
$\beta=-0.29$ \cite{sush,phdthesis} used for PbTiO$_3$ that was orginally derived for the garnet 
structure of a Gd$^{3+}$ ion enclosed by eight O$^{2-}$ ions \cite{sush1}.
Taking a value of $\beta\simeq -0.125$  for a Ti charge of $+3.5$ as an upper limit for $|\beta|$,
Eq. (\ref{deleps}) is evaluated for a Pb displacement of 0.47\AA, with effective principal
quantum numbers of the $6p$ and $7s$ states of Pb$^{2+}$, derived from Hartree-Fock
energies of 1.028 and 1.417 respectively \cite{effprin}. The nuclear radius, $r_0$, of $^{207}$Pb is 
taken $r_0\simeq CA^{1/3}\simeq 7.39 \times 10^{-15}$ m, where $A=207$ is the atomic mass number and  
$C = 1.25 \times 10^{-15}$ m. 
The resulting energy shift is then found to be,
\begin{equation}\label{enshift1}
\Delta\epsilon\simeq -0.66 \times 10^6\frac{({\bf X\cdot S})}{a_Bea_B^3}\,{\rm eV}
=-0.59 \times 10^6 \frac{S}{ea_B^3}\,{\rm eV}.
\end{equation}
This is approximately two times lower than the previous estimate, see Eq. (\ref{enshift}). 
We reiterate that expected uncertainty of this result is about 30-50\%,
see the discussion in Section 2.

\section{Conclusions}

The work in this paper is part of a continuing theoretical and experimental programme in aid of 
experimental investigations of CP violating effects in condensed matter systems. 
In the present work we have calculated the energy shift caused by
the nuclear Schiff moment of $^{207}$Pb in the ferroelectric PbTiO$_3$.
A Hartree-Fock
calculation of the PbO$_{12}$ cluster enabled the density asymmetry around the Pb nucleus due to 
the penetration of the oxygen electrons into the Pb ion to be calculated. From this an energy 
shift was calculated that is almost a factor of 2 lower than a previous estimate \cite{sush}.
It is recommended that the present estimate of the energy shift in Eq. (\ref{enshift1}) supercede 
the earlier result of \cite{sush}. 

There are two areas in the current work that merit further investigation. Firstly, the effect
of the potential due to the crystal field needs to be properly taken into account. A possible 
solution to this issue is via the CRYSTAL code \cite{crystal1,crystal2}. This particular code uses 
linear combinations of atom centred Gaussian functions 
to perform both Hartree-Fock and a variety of density functional theory calculations. This would allow 
a good description of both the crystal structure and also the penetration of the O electrons into the 
Pb ion. An existing calculation using the CRYSTAL code has provided 
accurate structural and electronic properties of PbTiO$_3$ \cite{bilc}. 
As we discussed above the DFT calculation would also allow an estimate to be made of the
magnitude of the correlation correction. Hopefully it would also allow an extension of the size of the
dynamic cluster.

In the present work we account for relativistic effects via effective renormalization
of non-relativistic matrix elements. According to previous calculations for heavy atoms this method
works pretty well. However, this has never been checked for solids.
Therefore the account of relativistic effects inside the Hartree-Fock or DFT method would
be desirable.  As a first step, this 
can be done by solving the Dirac-Fock equations for the PbO$_{12}$ cluster. A challenge here will be
in converging the self-consistent 4-component Dirac-Fock equations for such a large system. The BERTHA
\cite{bertha} or DIRAC codes \cite{dirac} can be investigated for this purpose.

These two phenomena that must be included in any complete treatment of the problem, namely the potential 
due to the crystal lattice and relativistic effects due to the high nuclear charge of the Pb nucleus, make 
this a highly demanding theoretical and computational problem.

A positive experimental finding of the nuclear Schiff moment would be of great importance to the 
entire physics community. Theoretical input into the design and interpretation of future experiments 
will be crucial in this effort. As such, we hope that the present model calculations will 
prompt renewed theoretical interest in this field.

\section*{Acknowledgements}

We are very grateful to H.~M.~Quiney, T.~N.~Mukhamedjanov and G.~F.~Gribakin for helpful
advice and discussions. This work was supported by the Australian Research Council.\\

%****************************************************************************

\begin{thebibliography}{99}
\bibitem{effprin} We follow Ref. \cite{sush1} in defining the effective principal
quantum numbers for the neutral atom. This explains the absence of the ion charge $Z_i=2$
in Eq. (\ref{deleps}). See also Ref. \cite{buhmann} for more details.
\end{thebibliography}
\end{document}